\documentclass[12pt,a4paper]{article}

\newtheorem{thm}{Theorem}

\newtheorem{defn}{Definition}

\usepackage{graphics} 
\usepackage{epsfig} 
\usepackage{amsmath} 
\usepackage{amssymb}  
\usepackage{cite}

\def\be{\begin{eqnarray}}
\def\ee{\end{eqnarray}}
\def\bee{\begin{eqnarray*}}
\def\eee{\end{eqnarray*}}

\def\la{\langle}
\def\ra{\rangle}

\def\tr{{\rm Tr}}
\def\ot{\otimes}
\def\bra{\langle}
\def\ket{\rangle}

\newcommand{\norm}[2]{\ensuremath{#1 \hskip-4pt \rightarrow \hskip-2pt #2}}

\def\bmx{\begin{matrix}}
\def\emx{\end{matrix}}

\title{Multiplicativity of superoperator norms for some entanglement breaking channels}

    \author{Christopher King\\
    \\
{\small      Departments of Mathematics and Physics} \\
{\small      Northeastern University} \\
{\small      Boston MA 02115}}

\begin{document}

     \maketitle

     \begin{abstract}
It is known that the minimal output entropy is additive for any product of entanglement breaking (EB) channels. The same is true for the Renyi entropy, where
additivity is equivalent to multiplicativity of the \norm{1}{q} norm for all $q \ge 1$. In this paper we consider the related question of multiplicativity of the
\norm{2}{q} norm for entanglement breaking channels. We prove that multiplicativity holds in this case for certain classes of EB channels, including both the CQ and QC channels.
\end{abstract}

\section{Introduction}
The mathematical problem of multiplicativity for completely positive maps on matrix algebras (aka superoperators) has arisen in several different contexts in quantum information theory. The original motivation arose from the question of whether the capacity of a quantum channel to transmit classical information is additive~\cite{BFS}. This question led to the additivity conjecture for minimal output entropy~\cite{KR1}, and the multiplicativity conjectures for output purity~\cite{AHW} (all now known to be false in general). A more recent application of multiplicativity appeared in the proof of the strong converse of the channel coding theorem \cite{WWY}, where multiplicativity of the \norm{1}{q}  norm with $q > 1$ was the key ingredient in proving the result for entanglement-breaking channels. Another recent example was the resurgence of interest in hypercontractive bounds and the logarithmic Sobolev inequality for quantum channel semigroups \cite{KT,M1}. Here the key quantity of interest is the \norm{2}{q} norm with $q > 2$, and its behavior in the limit where $q \downarrow 2$. Multiplicativity of this quantity is related to explicit formulas for the logarithmic Sobolev constant of product semigroups. Less is known about multiplicativity for the \norm{2}{q} norm than for the more widely studied \norm{1}{q} norm. In this paper we present some new results on \norm{2}{q} multiplicativity for several special classes of entanglement-breaking maps.

First we review some notation and basic definitions.
The norm of a matrix $A \in {\cal M}_d = {\cal M}(\mathbb{C}^{d \times d})$ is defined for $p \ge 1$ by
\be
\| A \|_p = \left( \tr | A |^p \right)^{1/p}
\ee
The \norm{p}{q} norm of an operator $L : {\cal M}_d \rightarrow {\cal M}_{d'}$ is defined by
\be
\| L \|_{p \rightarrow q} = \sup_{A \in {\cal M}_d} \frac{\| L(A) \|_{q}}{\| A \|_p}
\ee
For completely positive (CP) maps it has been shown that the norm is achieved on positive semidefinite matrices \cite{Au1,Wat}.
We will denote by ${\cal M}_d^+$ the positive semidefinite matrices in ${\cal M}_d$, so we have
\be
\Phi \, \mbox{is CP} \, \Rightarrow \| \Phi \|_{p \rightarrow q} = \sup_{A \in {\cal M}_d^+} \frac{\| \Phi(A) \|_{q}}{\| A \|_p}
\ee
Our objective in this paper is to analyze conditions for multiplicativity of the \norm{p}{q} norm for products of certain CP maps, and for certain values of $p$ and $q$. In all cases our results will apply to a product of the form $\Phi \ot \Omega$, where $\Phi$ is the map of interest and $\Omega$ is any other CP map. Note that $\| \Phi \ot \Omega \|_{p \rightarrow q} \ge \| \Phi \|_{p \rightarrow q} \, \| \Omega \|_{p \rightarrow q}$ is always true.
So we make the following definition.

\begin{defn}
We will say that the CP map $\Phi$ is fully $(\norm{p}{q})$-multiplicative (abbreviated as $FM(p,q)$) if for every CP map $\Omega$ we have
\be\label{def:FM}
\| \Phi \ot \Omega \|_{p \rightarrow q} = \| \Phi \|_{p \rightarrow q} \, \| \Omega \|_{p \rightarrow q}
\ee
\end{defn}
It follows that if $\Phi$ is $FM(p,q)$ then for any integer $n$
\be
\| \Phi^{\ot n} \|_{p \rightarrow q} = \| \Phi \|_{p \rightarrow q}^n
\ee

It is known that every CP map is $FM(p,q)$ for all $p \ge q \ge 1$ \cite{DJKR}. For $p < q$ multiplicativity does not hold in general, though it is known for some special cases. In particular
the identity map is known to be $FM(p,q)$ for every $1 \le p \le 2 \le q$ \cite{Wat},
and the same is true for every unital qubit channel \cite{Ki4}. Also any channel whose Choi-Jamiolkowski matrix is entrywise positive is known to be
$FM(2,2n)$ for all integers $n$~\cite{Ki5}.

Recall that an entanglement-breaking (EB) map $\Phi : {\cal M}_{d} \rightarrow {\cal M}_{d'}$ can be written in the form
\be\label{def:EB}
\Phi(A) = \sum_{k=1}^N \tr (A X_k) \, R_k
\ee
where $\{X_k \in {\cal M}_{d}^+\}$  play the role of a generalized measurement on the input, and $\{R_k \in {\cal M}_{d'}^+\}$ are the output states. In general we do not require that $\{X_k\}$ form a POVM or that $\{R_k\}$ have unit trace. However two special cases where these conditions do hold are of particular interest: when $\{X_k = | \psi_k \ra \la \psi_k |\}$ and $\{ | \psi_k \ra\}$ form an orthonormal basis of $\mathbb{C}^d$, the map is called CQ. When $\{R_k = | \phi_k \ra \la \phi_k |\}$ and $\{ | \phi_k \ra \}$ form an orthonormal basis of $\mathbb{C}^{d'}$, the map is called QC.

All EB maps are $FM(1,q)$ for all $q \ge 1$~\cite{Ki6}. Less is known about whether EB maps are $FM(p,q)$ for $1 < p < q$, and that will be the focus of our results here. To this end we will consider two special classes of EB maps for which we can prove new results.

\begin{thm}\label{thm:GCQ-GQC}
Suppose that the EB map $\Phi$ as defined in (\ref{def:EB}) either has entrywise positive input measurements 
\be\label{cond1}
\la i | X_k | j \ra \ge 0 \, \mbox{ for all $i,j \in \{1,\dots,d\}$, all $k=1,\dots,N$}
\ee
or has diagonal output states
\be\label{cond2}
\la i | R_k | j \ra = 0 \, \mbox{ for all $i \neq j \in \{1,\dots,d'\}$, all $k=1,\dots,N$.}
\ee
Then $\Phi$ is $FM(2,q)$ for all $q > 2$.
\end{thm}

\medskip

\par\noindent
\noindent{\em Remark 1:} It is reasonable to conjecture that all EB maps are $FM(2,q)$ for all $q > 2$.

\medskip
\par\noindent
\noindent{\em Remark 2:} For any map $L$ on matrix algebras the adjoint is written $\widehat{L}$ and is defined by
\be
\tr B^* \, L(A) = \tr \left( \widehat{L}(B) \right)^* \, A
\ee
It follows that for all $p,q$ we have
\be
\| L \|_{p \rightarrow q} = \| \widehat{L} \|_{q' \rightarrow p'}
\ee
where $q',p'$ are the conjugate values for $q,p$. Thus a CP map $\Phi$ is $FM(p,q)$
if and only if $\widehat{\Phi}$ is $FM(q',p')$.
For an EB map $\Phi(A) = \sum_{k=1}^N \tr (A X_k) \, R_k$ we find
\be
\widehat{\Phi}(B) = \sum_{k=1}^N \tr (B R_k) \, X_k
\ee
Thus Theorem \ref{thm:GCQ-GQC} also implies that if $\Phi$ is an EB map of the form 
(\ref{def:EB}) satisfying (\ref{cond1}) and (\ref{cond2}) with the roles
of $X_k$ and $R_k$ switched, then $\Phi$ is $FM(p,2)$ for all $1 \le p < 2$.

\medskip
\par\noindent
\noindent{\em Remark 3:}  
One major motivation for considering the class $FM(2,q)$ is its relation to hypercontractivity, which we explain now.
It is convenient to introduce a new norm on matrices
\be\label{def:trip-norm}
||| A |||_p = \frac{|| A ||_p}{|| I_d ||_p}
\ee
and use it to define norms for maps:
\be
||| L |||_{p \rightarrow q} = \sup_{A} \frac{||| L(A) |||_q}{||| A |||_p}
\ee
If $||| L |||_{p \rightarrow q} \le 1$ we say that $L$ is a contraction from $L^p$ to $L^q$.
To see the motivation for this definition, note that the classical discrete case is recovered by the restriction to 
diagonal matrices, and in this case (\ref{def:trip-norm}) is computed using the uniform probability measure
on $\{1,\dots,d\}$: for a diagonal matrix $A = {\rm diag}(a_1,\dots,a_d)$ we have
\be
||| A |||_p = \left( \frac{1}{d} \sum_{i=1}^d |a_i|^p \right)^{1/p}
\ee
Then if the operator $L$ maps diagonal matrices to diagonal matrices and satisfies $||| L |||_{p \rightarrow q} \le 1$,
it is a contraction with respect to the uniform probability measure on its domain and range.

\medskip
Now suppose that $\Phi_t$ is a one-parameter semigroup of CP maps, with $\Phi_0 = {\cal I}$ the identity map, and 
satisfying
\be
\Phi_t(I_d) = I_d, \quad
\lim_{t \rightarrow \infty} \Phi_t(A) = \frac{1}{d} \tr (A) I_d \quad \mbox{all $A \in {\cal M}_d$}
\ee
We always have 
\be\label{norm-bd1}
||| \Phi_t |||_{p \rightarrow q} \ge \frac{||| \Phi_t(I_d) |||_q}{||| I_d |||_p} = \frac{||| I_d |||_q}{||| I_d |||_p} = 1
\ee
We say the semigroup $\Phi_t$ satisfies a {\em hypercontractivity bound} for $(p,q)$ if there is $t(p,q) \ge 0$ such that
\be
||| \Phi_t |||_{p \rightarrow q} = 1 \quad \mbox{if and only if} \quad t \ge t(p,q)
\ee
If the equation $t = t(2,q)$ can be solved for $q=q(t)$ in some interval $0 \le t \le T$ with $q(0)=2$ then we have
\be
||| \Phi_t |||_{2 \rightarrow q(t)} = 1, \quad\quad 0 \le t \le T
\ee
Taking the derivative at $t=0$ leads to the logarithmic Sobolev inequality.
Various cases of this have been proven for different classes of semigroups. 
One important example is the depolarizing channel
$\Delta_{\lambda}(\rho) = \lambda \rho + (1-\lambda)/d \,\, \tr (\rho) \, I_d$, where $\lambda = e^{-t}$.
For the depolarizing channel the value of
$t(p,q)$ is known explicitly, and also the log Sobolev constant~\cite{KT}.

\medskip
The relation to multiplicativity arises when considering a product of semigroups.
If multiplicativity holds then
\be
|| \Phi_t \ot \Phi_t ||_{p \rightarrow q} = || \Phi_t ||_{p \rightarrow q}^2
\ee
The normalization factors are the same on both sides so this also implies
\be
||| \Phi_t \ot \Phi_t |||_{p \rightarrow q} = ||| \Phi_t |||_{p \rightarrow q}^2
\ee
Thus multiplicativity implies that the time to contractivity $t(p,q)$ is the same for $\Phi_t$ and for the product
$\Phi_t \ot \Phi_t$. This allows explicit computation of the logarithmic Sobolev constant for the product map.

\section{Proof of Theorem \ref{thm:GCQ-GQC}}
We will first prove the case where the matrices $X_k$ are entrywise positive.
Let $\Phi$ be an EB map satisfying (\ref{cond1}), and let $\Omega$ be any CP map. We
will be concerned with the output matrix $(\Phi \ot \Omega)(\rho)$ where $\rho$ is any bipartite
input matrix satisfying $\| \rho \|_2 =1$:
\be\label{in1}
\rho = \sum_{i,j} | i \ket \bra j | \ot \rho_{ij}, \qquad
\sum_{i,j} \| \rho_{ij} \|_2^2 = 1
\ee
The output can be written
\be\label{out1}
(\Phi \ot \Omega)(\rho) = \sum_{i,j} \sum_{k=1}^N \tr (| i \ket \bra j | X_k) \, R_k  \ot \Omega(\rho_{ij}) = 
\sum_{k} R_k \ot \Omega(A_k)
\ee
where we have defined 
\be\label{def:Ak}
A_k = \sum_{i,j} \bra j | X_k | i \ket \, \rho_{ij} = \tr_1 \left(X_k \ot I\right) \rho
\ee
Since $X_k \ge 0$ and $\rho \ge 0$, it follows that $A_k \ge 0$ for all $k$.
We define for all $i,j = 1,\dots,d$
\be\label{def:tau}
\tau_{ij} = \| \rho_{ij} \|_2
\ee
and let $\tau \in {\cal M}_d$ denote the corresponding matrix. It is easy to check that
\be\label{tau-norm}
\| \tau \|_2^2 = \sum_{ij}  \, \tau_{ij}^2 = \sum_{ij} \tr | \rho_{ij} |^2 = \| \rho \|_2^2 = 1
\ee
Furthermore
\be\label{ineq4}
\| A_k \|_2^2
&=& \bigg| \sum_{i,j,m,n} \overline{\la j | X_k | i \ra} \, \la n | X_k | m \ra \, \tr (\rho_{ij}^* \rho_{mn}) \bigg| \nonumber \\
& \le & \sum_{i,j,m,n} | \overline{\la j | X_k | i \ra}| \, |\la n | X_k | m \ra | \, \tau_{ij} \, \tau_{mn} \nonumber \\
&=&  \sum_{i,j,m,n} \la j | X_k | i \ra \, \la n | X_k | m \ra \, \tau_{ij} \, \tau_{mn} \nonumber \\
&=& ( \tr (X_k \tau) )^2
\ee
where we used the entrywise positivity of $X_k$, and also
the Cauchy-Schwarz inequality to deduce that
\be
| \tr (\rho_{ij}^* \rho_{mn}) | \le \tau_{ij} \, \tau_{mn}
\ee
We also define for each $k=1,\dots,N$
\be
\theta_k = ( \tr (X_k \tau) )^{-1} \, A_k
\ee
so that (\ref{out1}) becomes
\be\label{out2}
(\Phi \ot \Omega)(\rho)  =
\sum_{k} \tr (X_k \tau) \, R_k \ot \Omega(\theta_k)
\ee
Since both $X_k$ and $\tau$ are entrywise positive, we have $\tr (X_k \tau) \ge 0$, and from
(\ref{def:Ak}) we have $A_k \ge 0$. Therefore $\theta_k \ge 0$, and
from the bound (\ref{ineq4}) we get
\be\label{theta-norm}
\| \theta_k \|_2 \le 1
\ee

\medskip
We are now ready to prove the multiplicativity bound. 
The proof will use the Lieb-Thirring matrix inequality~\cite{LiebTh}: for all positive matrices $C$, all matrices $B$ and all $q \ge 1$,
\be\label{L-T}
\tr (B^* C B)^q \le \tr (B B^*)^q C^q
\ee
To apply this bound we first define
\be\label{def:zk}
z_k = (\tr (X_k \tau) \, R_k)^{1/2}, \quad k=1,\dots,N
\ee
Note that
\be
\sum_k z_k^2 = \sum_k \tr (X_k \tau) \, R_k = \Phi(\tau)
\ee
We can  rewrite the output matrix (\ref{out2}) as follows:
\be\label{out3}
(\Phi \ot \Omega)(\rho)  &=& \sum_{k} \tr (X_k \tau) \, R_k \ot \Omega(\theta_k) \nonumber \\
&=& \sum_{k} z_k^2 \ot \Omega(\theta_k) \nonumber \\
&=&  \sum_{k} (z_k \ot I) \, (I \ot \Omega(\theta_k)) \, (z_k \ot I) \nonumber \\
&=&  \sum_{k} ( \bra k | \ot z_k \ot I ) \, (| k \ket \bra k | \ot I \ot \Omega(\theta_k)) \, ( | k \ket \ot z_k \ot I ) \nonumber \\
&=&  \sum_{i,j,k} ( \bra i | \ot z_i \ot I ) \, (| j \ket \bra j | \ot I \ot \Omega(\theta_j)) \, ( | k \ket \ot z_k \ot I ) \nonumber \\
&=& B^* C B
\ee
We have introduced a third space in the tensor product above without changing the value of the output. 
We have also introduced the matrices
\be
B &=& \sum_{k} | k \ket \ot z_k \ot I, \\
C &=& \sum_j | j \ket \bra j | \ot I \ot \Omega(\theta_j) 
\ee
Since each matrix $\theta_j$ is positive and $\Omega$ is CP, 
it follows that $C$ is positive, and hence we can apply the Lieb-Thirring inequality (\ref{L-T}). 
We will write $\tr_{12}$ to denote trace over the first two factors etc.
The right side of (\ref{L-T}) is
\be\label{L-T2}
\tr (B B^*)^q C^q &=& \tr_{123} \left(\sum_{i,k} | k \ket \bra i | \ot z_k z_i \ot I \right)^q \, \left( \sum_j | j \ket \bra j | \ot I \ot \Omega(\theta_j)\right)^q \nonumber \\
&=& \tr_{123} \left(\sum_{i,k} | k \ket \bra i | \ot z_k z_i \ot I \right)^q \, \left( \sum_j | j \ket \bra j | \ot I \ot \Omega(\theta_j)^q \right) \nonumber \\
&=& \sum_j \tr_{123} \left[ \left(\sum_{i,k} | k \ket \bra i | \ot z_k z_i \right)^q \ot I \right] \, \bigg[ | j \ket \bra j | \ot I \ot \Omega(\theta_j)^q \bigg] \nonumber \\
&=& \sum_j \tr_{12} \left[\left(\sum_{i,k} | k \ket \bra i | \ot z_k z_i \right)^q  \, \bigg( | j \ket \bra j | \ot I \bigg) \right] \tr_{3} \bigg[\Omega(\theta_j)^q \bigg] \nonumber \ee
By definition of the \norm{2}{q} norm and using (\ref{theta-norm}) we have
\be
\tr_{3} \bigg[\Omega(\theta_j)^q \bigg] \le \| \Omega \|_{2 \rightarrow q}^q \, \| \theta_j \|_2^q = \| \Omega \|_{2 \rightarrow q}^q 
\ee
Thus we get the bound
\be\label{L-T3}
\tr (B B^*)^q C^q
&\le& \| \Omega \|_{2 \rightarrow q}^q \,  \sum_j \tr_{12} \left[\left(\sum_{i,k} | k \ket \bra i | \ot z_k z_i \right)^q  \, \bigg( | j \ket \bra j | \ot I \bigg) \right]  \nonumber \\
& = & \| \Omega \|_{2 \rightarrow q}^q \, \tr_{12} \left(\sum_{i,k} | k \ket \bra i | \ot z_k z_i \right)^q  \nonumber \\
& = & \| \Omega \|_{2 \rightarrow q}^q \, \tr_{12} \left(\left[\sum_{k} | k \ket  \ot z_k\right] \,  \left[\sum_{i}  \bra i | \ot z_i\right]\right)^q  \nonumber \\
& = & \| \Omega \|_{2 \rightarrow q}^q \, \tr_{12} \left( \left[\sum_{i}  \bra i | \ot z_i\right] \, \left[\sum_{k} | k \ket  \ot z_k\right]\right)^q  \nonumber \\
& = & \| \Omega \|_{2 \rightarrow q}^q \, \tr_{2} \left( \sum_{k}  z_k^2 \right)^q  \nonumber \\
& = & \| \Omega \|_{2 \rightarrow q}^q \, \tr_{2} \left( \Phi(\tau) \right)^q  \nonumber \\
&\le& \| \Omega \|_{2 \rightarrow q}^q \, \| \Phi \|_{2 \rightarrow q}^q \, \| \tau \|_2^q \nonumber \\
&=& \| \Omega \|_{2 \rightarrow q}^q \, \| \Phi \|_{2 \rightarrow q}^q
\ee
where we used (\ref{tau-norm}) in the last step.

\medskip
Combining (\ref{out3}) and (\ref{L-T3}) we deduce that $\Phi$ is $FM(2,q)$: for all $q \ge 2$
\be
\| \Phi \ot \Omega \|_{2 \rightarrow q} \le \| \Phi \|_{2 \rightarrow q} \, \| \Omega \|_{2 \rightarrow q}
\ee

\medskip
For the proof of the second part of Theorem \ref{thm:GCQ-GQC} we assume that the output matrices $R_k$ are diagonal
(with non-negative entries). We first rewrite (\ref{in1}) as follows:
\be\label{in2}
\rho = \sum_{i,j} \tilde{\rho}_{ij} \ot | i \ket \bra j |, \qquad
\sum_{i,j} \| \tilde{\rho_{ij}} \|_2^2 = 1
\ee
This leads to the relations
\be\label{out1a}
(\Phi \ot \Omega)(\rho) = \sum_{k} R_k \ot \Omega(B_k)
\ee
where we have defined 
\be\label{def:Bk}
B_k = \sum_{i,j} \tr \left( \tilde{\rho}_{ij} X_k \right)  \, | i \ket \bra j |
\ee
Since $R_k$ is diagonal we can write
\be
R_k = \sum_m r_{km} \, | m \ket \bra m |, \qquad r_{km} \ge 0
\ee
and then
\be\label{out2a}
(\Phi \ot \Omega)(\rho) = \sum_{m}  | m \ket \bra m | \ot \Omega(C_m), \qquad
C_m = \sum_k r_{km} \, B_k
\ee
This leads immediately to
\be\label{out3a}
\tr (\Phi \ot \Omega)(\rho)^q = \sum_m \tr \, \Omega(C_m)^q \le \| \Omega \|_{2 \rightarrow q}^q \,\sum_m \|C_m\|_2^q
\ee
Now we define the normalized states
\be
\sigma_{ij} = \frac{1}{\| \tilde{\rho}_{ij} \|_2} \, \tilde{\rho}_{ij}, \qquad
\| \sigma_{ij} \|_2 = 1
\ee
and also 
\be
Y_m = \sum_k r_{km} \, X_k
\ee
Inserting into (\ref{def:Bk}) we find
\be 
C_m = \sum_{i,j} \tr \left( \tilde{\rho}_{ij} Y_m \right)  \, | i \ket \bra j | 
\ee
It follows that
\be
\|C_m\|_2^2 = \sum_{ij} \bigg|  \tr (Y_m \tilde{\rho}_{ij}) \bigg|^2 =
\sum_{ij} \| \tilde{\rho}_{ij} \|_2^2 \, \bigg|  \tr (Y_m \sigma_{ij}) \bigg|^2
\ee
Since $q \ge 2$, the map $x \rightarrow x^{q/2}$ is convex, and hence using (\ref{in2}) we get
\be
\|C_m\|_2^q &=& \left( \sum_{ij} \| \tilde{\rho}_{ij} \|_2^2 \, \bigg|  \tr (Y_m \sigma_{ij}) \bigg|^2 \right)^{q/2} \nonumber \\
& \le & \sum_{ij} \| \tilde{\rho}_{ij} \|_2^2 \, \bigg|  \tr (Y_m \sigma_{ij}) \bigg|^q
\ee
Combining with (\ref{out3a}) this gives
\be\label{out4a}
\tr (\Phi \ot \Omega)(\rho)^q & \le & \| \Omega \|_{2 \rightarrow q}^q \, \sum_m \, \sum_{ij} \| \tilde{\rho}_{ij} \|_2^2 \, \bigg|  \tr (Y_m \sigma_{ij}) \bigg|^q \nonumber \\
& = & \| \Omega \|_{2 \rightarrow q}^q \, \sum_{ij} \| \tilde{\rho}_{ij} \|_2^2 \,\sum_m \, \bigg|  \tr (Y_m \sigma_{ij}) \bigg|^q
\ee
Finally we notice that
\be
\Phi(\sigma_{ij}) &=& \sum_k R_k \, \tr \left(X_k \sigma_{ij} \right) \nonumber \\
&=& \sum_m | m \ket \bra m | \, \tr \left(Y_m \sigma_{ij} \right)
\ee
and therefore
\be
\tr | \Phi(\sigma_{ij})|^q = \sum_m \, \bigg|  \tr (Y_m \sigma_{ij}) \bigg|^q
\ee
So finally from (\ref{out4a}) we get
\be\label{out5a}
\tr (\Phi \ot \Omega)(\rho)^q & \le & \| \Omega \|_{2 \rightarrow q}^q \, \sum_{ij} \| \tilde{\rho}_{ij} \|_2^2 \,\tr | \Phi(\sigma_{ij})|^q \nonumber \\
& \le & \| \Omega \|_{2 \rightarrow q}^q \, \| \Phi \|_{2 \rightarrow q}^q \,\sum_{ij} \| \tilde{\rho}_{ij} \|_2^2 \, \| \sigma_{ij} \|_2^q \nonumber \\
& =&\| \Omega \|_{2 \rightarrow q}^q \, \| \Phi \|_{2 \rightarrow q}^q \,\sum_{ij} \| \tilde{\rho}_{ij} \|_2^2 \nonumber \\
&=& \| \Omega \|_{2 \rightarrow q}^q \, \| \Phi \|_{2 \rightarrow q}^q
\ee
which completes the proof.

{~~}

\end{document}